\documentclass[12pt]{article}

\usepackage[utf8]{inputenc}
\usepackage[T1]{fontenc}
\usepackage{times}                  
\usepackage[
  letterpaper,
  top=1in, bottom=1in,
  left=1in, right=1in
]{geometry}
\usepackage{amsmath, amssymb, amsfonts}
\usepackage{graphicx}
\usepackage{booktabs}
\usepackage{siunitx}
\usepackage{multirow}
\usepackage{setspace}
\usepackage{parskip}
\usepackage{titlesec}
\usepackage{caption}
\usepackage{hyperref}
\usepackage[numbers,sort&compress]{natbib}
\usepackage[table]{xcolor}
\titleformat{\section}{\normalfont\bfseries\normalsize}{\thesection.}{0.5em}{}
\titleformat{\subsection}{\normalfont\bfseries\normalsize}{\thesubsection}{0.5em}{}
\titleformat{\subsubsection}{\normalfont\bfseries\normalsize}{\thesubsubsection}{0.5em}{}
\titlespacing*{\section}{0pt}{12pt}{6pt}
\titlespacing*{\subsection}{0pt}{10pt}{4pt}
\titlespacing*{\subsubsection}{0pt}{8pt}{2pt}

\date{}

\begin{document}

\begin{center}
{\large\bfseries
Joint Forecasting of Extreme Events through Dual-Stage Cascade\\
Reservoir Computing}
\end{center}

\vspace{0.6em}

\begin{center}
Yueyang Wang,$^{1}$\;
Juncheng Huang,$^{1}$\;
Hanxu Zhou,$^{1}$\;
Tao Wang$^{1,*}$
\end{center}

\vspace{0.4em}

\begin{center}
\textit{%
$^{1}$State Key Laboratory of Integrated Service Networks,
School of Communications Engineering, Xidian University,
Xi'an, Shaanxi 710071, China}
\end{center}

\vspace{0.3em}

\begin{center}
\textbf{Email:} \textit{twang6@xidian.edu.cn (T. Wang)}
\end{center}

\vspace{1em}

\noindent\textbf{Abstract}

\noindent

Reservoir computing (RC) offers an efficient data-driven approach for forecasting extreme events (EEs), which correspond to rare and large-amplitude dynamical occurrences. We propose a dual-stage cascade framework that jointly predicts both the timing and peak intensity of upcoming EEs.A traditional RC branch integrates long-term precursor dynamics to support
stable detection and long-horizon prediction, while an NGRC branch captures
local nonlinear waveform geometry to improve fine-grained time-to-peak
localization and complement peak-intensity estimation. The fused features then feed a ridge classifier that issues a binary alarm upon detecting precursors. Only then do two ridge regressors, trained on true-positive snapshots, estimate time-to-peak and peak intensity. This classify-then-regress design addresses severe class imbalance without data resampling. Evaluated on simulated pump-modulated VCSEL data, the hybrid model achieves a SEDI value $>0.8$, with MAEs around 0.2~ns and 0.2~a.u., maintaining performance up to a 20~ns warning horizon. The framework advances extreme-event forecasting from binary warnings to fully quantitative dual-objective prediction.

\vspace{0.8em}


\vspace{0.8em}

\noindent\textit{Innovative Description:}
A dual-stage cascade reservoir computing framework jointly predicts
extreme optical event timing and intensity.

\vspace{1em}

\section{Introduction}

Extreme events (EEs), featured by rare, large-amplitude excursions that far
exceed typical fluctuations of a system, have been documented across an
extraordinary range of natural and laboratory settings. These span from oceanic
rogue waves~\cite{Dysthe2008, Pelinovsky2008, Osborne2010} to astrophysical
transients~\cite{Hudson1991} to atmospheric anomalies~\cite{Stenflo2009} and
optical instabilities~\cite{Metayer2014, Baronio2012, Dudley2009, Bonatto2011,
Onorato2013, Akhmediev2016}. Despite their diversity, these phenomena share two
common features: they emerge from nonlinear dynamics often corrupted by noise,
and their consequences are disproportionately
severe~\cite{Chowdhury2022, Farazmand2019, Ummenhofer2017, Raymond2020}. While
reliable early warning, even if only a few characteristic timescales in
advance, could substantially reduce damage, EEs remain notoriously difficult
to forecast. In many systems, such as semiconductor lasers, the intrinsic
stochasticity of spontaneous emission noise renders the exact timing and
amplitude of these bursts fundamentally unpredictable from deterministic models
alone.

This inherent unpredictability has motivated a significant shift toward
model-free, data-driven forecasting. Among various machine learning algorithms,
RC has attracted particular attention for its ability to project complex
nonlinear dynamics into a high-dimensional state space while maintaining a
computationally efficient, closed-form training
procedure~\cite{Pathak2018, Ghosh2021, Tanaka2019, Jaeger2001, Jaeger2004,
Maass2002, Lukosevicius2009}. RC has demonstrated remarkable success in chaotic
attractor reconstruction~\cite{Racca2021, Vlachas2020}, climate
prediction~\cite{Pei2023, Nadiga2021}, and epidemic
forecasting~\cite{Ghosh2021, Saha2022, Carroll2018}. Within the specific
context of extreme events, prior studies have shown that RC can issue binary
alarms shortly before a burst~\cite{Suda2020, Wang2024} or estimate the
long-term statistical properties of event amplitudes~\cite{Racca2021}.
However, a unified framework capable of simultaneously forecasting both the
arrival time and the peak intensity of an impending event has remained elusive.
This challenge arises because reliable event detection requires the integration of precursor information
over an extended temporal context, whereas accurate time-to-peak localization
benefits from resolving the local nonlinear geometry of the evolving waveform.
Peak-intensity estimation depends on both sources of information: the
longer-term dynamical state constrains the overall burst magnitude, while local
waveform features provide complementary information about its subsequent
development of the waveform geometry.

\begin{figure}[t]
  \centering
  \includegraphics[width=0.9\textwidth, height=0.5\textheight]{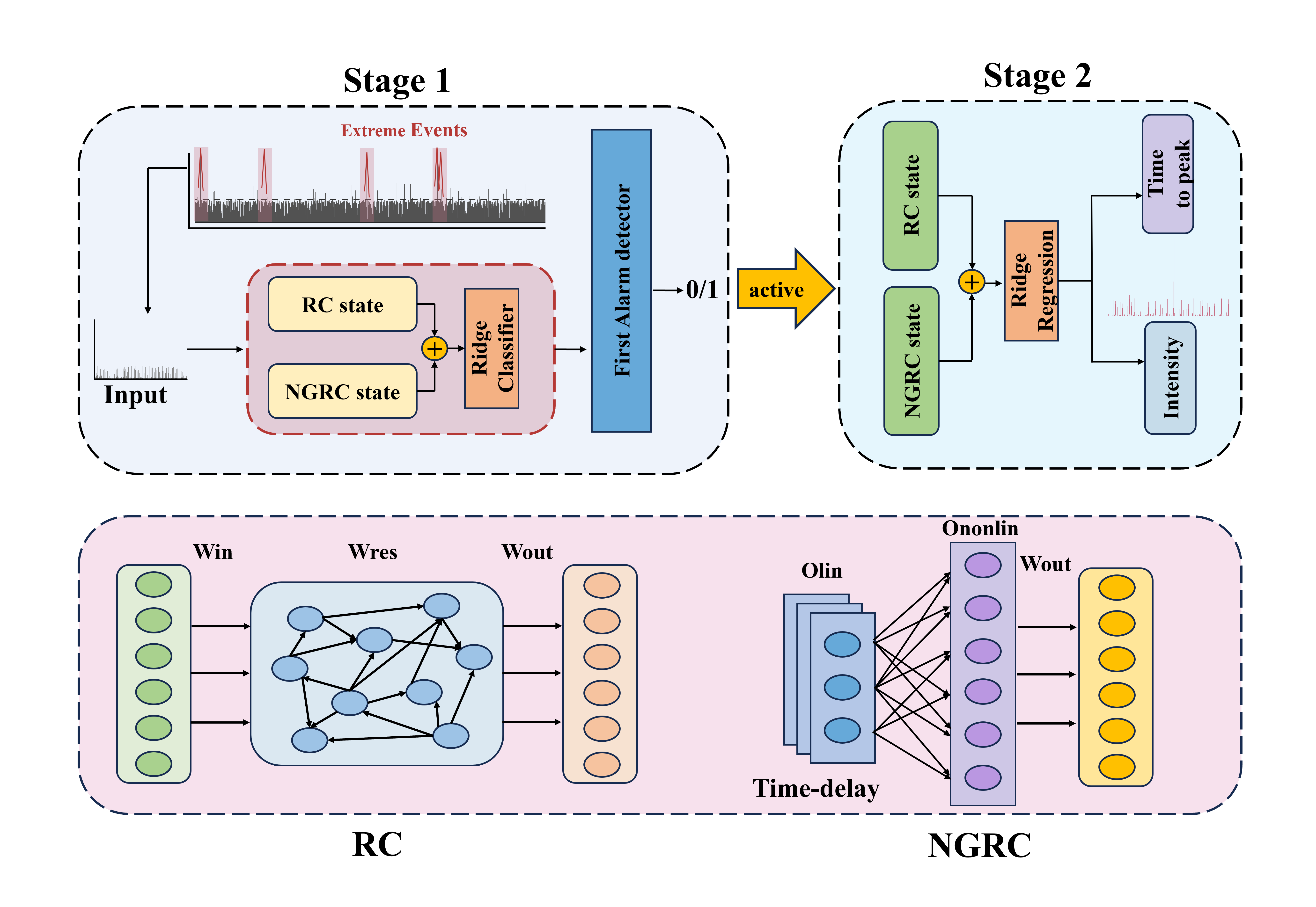}
  \caption{\textbf{Architecture of the dual-stage alarm-triggered cascade
    framework.}
    (Left, Stage~1) The scalar input $x_t$ is processed by an RC branch
    (echo state network) and an NGRC branch (polynomial delay embedding) in
    parallel. Their outputs are concatenated into a fused feature vector
    $\mathbf{f}_t$ that feeds a ridge classifier, which issues a binary alarm
    (0/1) at the first alarm time $t_{\mathrm{fa}}$. (Right, Stage~2)
    Upon alarm, $\mathbf{f}_{t_{\mathrm{fa}}}$ is passed to two independent
    ridge regressors that simultaneously predict the time-to-peak $\Delta t$
    and peak intensity $I^*$.}
  \label{fig:arch}
\end{figure}

In this work, we propose a dual-stage, warning-triggered cascade framework
designed to jointly forecast these two critical quantities. As shown in
Fig.~\ref{fig:arch}, our architecture fuses two complementary feature
extractors: a conventional RC branch that accumulates long-range temporal
memory, and a NGRC branch that captures instantaneous, high-order geometric
features of the waveform~\cite{Gauthier2021, Chepuri2024, Barbosa2022,
Bollt2021}. The concatenated output of these branches feeds a ridge classifier
that issues a binary alarm upon detecting precursor signatures. Only when an
alarm is triggered does the second stage engage: two independent ridge
regressors, trained exclusively on true-positive snapshots, simultaneously
estimate the time-to-peak and the peak intensity. This ``classify-then-regress''
design addresses the challenge of class imbalance, where extreme events
typically occupy less than 1\% of the data, by converting the signal into a
near-balanced regression task without the need for artificial resampling.

We validate this framework using the data from a pump-modulated
vertical-cavity surface-emitting laser (VCSEL), a system known for its complex
polarization-driven extreme events~\cite{Wang2024, wang2025extreme}. Across
five independent random seeds, the hybrid model achieves a SEDI exceeding 0.8.
The system yields mean absolute errors (MAE) of approximately 0.2~ns for
time-to-peak and 0.2~a.u.\ for peak intensity, even when the warning window is
shifted up to 20~ns ahead of the threshold crossing---a lead time corresponding
to roughly 125 characteristic autocorrelation times. Ablation experiments
confirm a clear functional dissociation: the RC branch is essential for stable
detection, while the NGRC branch is critical for timing precision. By bridging
the gap between binary early warning and fully quantitative prediction, this
work provides a robust pipeline for the real-time forecasting of rare,
high-impact excursions in complex dynamical systems.

\section{Methods and Materials}

\subsection{Framework Overview}

Fig.~\ref{fig:arch} summarizes the proposed dual-stage alarm-triggered
prediction framework. The system receives a scalar polarization intensity time
series and processes it through two parallel feature-extraction branches: a RC
branch and a NGRC branch. Their outputs are concatenated into a single fused
feature vector that drives a two-stage decision pipeline. In Stage~1, a ridge
classifier continuously monitors the fused features and raises a binary alarm
at the first alarm time $t_{\mathrm{fa}}$ once precursor signatures are
detected within a warning window preceding each extreme event. Stage~2
activates only upon alarm, at which point two independent ridge regressors
estimate the time-to-peak $\Delta t$ and peak intensity $I^*$ of the impending
event.

Conventional approaches that attempt to predict extreme events directly from
continuous time series inevitably suffer from severe class imbalance, because
the overwhelming majority of training samples represent quiescent
background~\cite{Wang2024}. Our classify-then-regress design circumvents this
by relegating the regression tasks to the sparse subset of true-positive alarm
instances identified by Stage~1. The Stage~2 regressors therefore specialize
entirely on the nonlinear transient dynamics immediately preceding the burst.
All learnable components are trained via closed-form ridge optimization with no
gradient computation required.

\begin{figure}[t]
\centering
\begin{minipage}{0.48\textwidth}
  \centering
  \includegraphics[width=\linewidth]{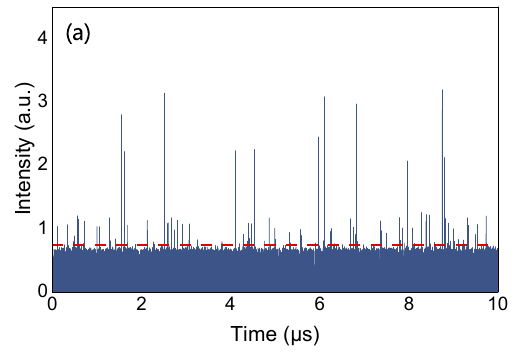}
\end{minipage}
\hfill
\begin{minipage}{0.49\textwidth}
  \centering
  \raisebox{0.25cm}{\includegraphics[width=\linewidth]{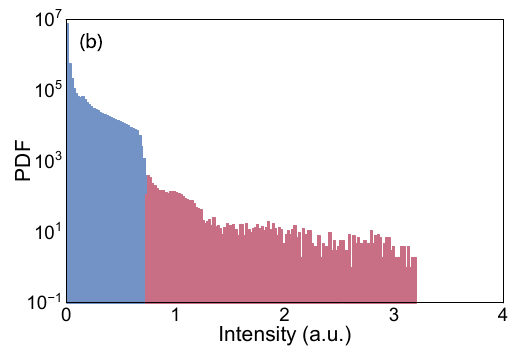}}
\end{minipage}
\vspace*{1.0em}
\caption{\textbf{Simulated polarization intensity of the VCSEL under pump
modulation.} (a) A representative segment of the raw intensity trace
$x_t$; the red dashed line marks the EE threshold $\theta$.
(b) Sorted intensity profile: the blue region corresponds to the
sub-threshold background, while the red tail represents
threshold-exceeding extreme events, confirming the heavy-tailed
statistical character of the emission.}
\label{fig:data}
\end{figure}

\subsection{Spin-Flip Model and Numerical Simulation}

The polarization dynamics of the VCSEL are governed by the spin-flip model
(SFM)~\cite{SanMiguel1995, MartinRegalado1997}, which describes the slowly varying
complex amplitudes $E_{\pm}$ of the right- and left-circularly polarized modes,
the total normalized carrier density $N$, and the carrier spin difference $n$:
\begin{align}
\frac{dE_{\pm}}{dt} &= \kappa(1+i\alpha)\bigl[(N \pm n) - 1\bigr]E_{\pm}
  - (\gamma_a + i\gamma_p)E_{\mp} \nonumber\\
&\quad + \sqrt{\beta(N \pm n)}\xi_{\pm}, \label{eq:sfm_field}\\[4pt]
\frac{dn}{dt} &= -\gamma_s n
  - \gamma\bigl[(N+n)|E_+|^2 - (N-n)|E_-|^2\bigr], \label{eq:sfm_n}\\[4pt]
\frac{dN}{dt} &= -\gamma\bigl[N - I(t)
  - (N+n)|E_+|^2 + (N-n)|E_-|^2\bigr]. \label{eq:sfm_N}
\end{align}
Here $\kappa$ is the field decay rate, $\alpha$ the linewidth enhancement
factor, $\gamma$ the carrier decay rate, and $\gamma_s$ the spin-flip
relaxation rate. Linear dichroism and birefringence are represented by
$\gamma_a$ and $\gamma_p$, respectively. The terms $\xi_{\pm}$ are independent
complex Gaussian Langevin noise sources with zero mean and correlation
$\langle \xi_\pm(t),\xi_\pm^*(t')\rangle = 2\delta(t-t')$, scaled by the
spontaneous emission coefficient $\beta$.

The normalized injection current is sinusoidally modulated to reproduce the
pump conditions under which extreme events arise~\cite{Wang2024}:
\begin{equation}
I(t) = I_{dc} + \Delta I \cos(2\pi f_m t),
\label{eq:pump}
\end{equation}
where $I_{dc}$ is the DC bias, $\Delta I$ the modulation amplitude, and $f_m$
the modulation frequency. All current parameters are expressed in dimensionless
normalized units.

The experimentally observable quantity used throughout this work is the linear
$x$-polarization intensity $x(t) = \tfrac{1}{2}|E_+ + E_-|^2$. Extreme events
emerge from the interplay between birefringence and spin relaxation. The
Langevin noise terms provide stochastic kicks that make individual event times
inherently unpredictable, even though the underlying attractor is deterministic.
This is precisely what motivates a data-driven warning strategy rather than a
closed-form analytical approach.

The SFM equations form a system of stochastic differential equations with
additive complex Gaussian noise. We integrate them with a stochastic
Runge--Kutta scheme of weak order two~\cite{Honeycutt1992, KloedenPlaten1992}
at fixed step size $\Delta t_i$, which preserves the correlation structure of
the Langevin forcing and reduces to the standard fourth-order deterministic
Runge--Kutta method in the limit $\beta \to 0$. The resulting intensity time
series $x_t$ is then down-sampled at interval $\Delta t_s$. All normalization
statistics are computed exclusively on the training partition to prevent
information leakage into validation or test splits.

\subsection{Extreme Event Definition and Labeling}
\label{sec:ee_def}

An extreme event is declared at the first index $t_0$ where the intensity
crosses the threshold~\cite{Bonatto2011}:
\begin{equation}
\theta = \mu_{\mathrm{tr}} + k\sigma_{\mathrm{tr}},
\label{eq:threshold}
\end{equation}
with $\mu_{\mathrm{tr}}$ and $\sigma_{\mathrm{tr}}$ being the mean and standard
deviation of the training set, and $k$ the threshold multiplier. Following the standard criterion for optical extreme events established by Bonatto et al.~\cite{Bonatto2011}, which defines a rogue wave pulse as exceeding the mean by eight standard deviations, and consistent with prior work on this VCSEL model~\cite{Wang2024}, we set $k = 8.0$. This ensures that fewer
than 1\% of time-series points exceed the threshold. Once a crossing is
identified, the peak index $t^*$ is located within a subsequent search window
$W_p$.

To evaluate detection and prediction at different lead times, we introduce a
non-negative shift parameter $\Delta W$ (in units of sampling steps). The
binary classifier label is set to $y_t = 1$ for
$t \in [t_0 - W - \Delta W, t_0 - \Delta W]$, where $W$ is the warning window
length (in steps) and $t_0$ is the threshold-crossing index. Equivalently,
this means the warning window of length $W$ is positioned such that its
rightmost edge is $\Delta W$ steps before $t_0$. Thus, a larger $\Delta W$
corresponds to an earlier warning window and a longer minimum advance notice
before the threshold crossing. The first alarm fires at the earliest time step
$t_{\mathrm{fa}}$ within this shifted window where the classifier score exceeds
the decision threshold. Upon alarm, the two regression targets are the
time-to-peak $\Delta t = (t^* - t_{\mathrm{fa}})\Delta t_s$ and the peak
intensity $I^* = x_{t^*}$. Note that the actual prediction horizon is
$\Delta t$, which inherently includes $\Delta W$ plus the interval from alarm
to $t_0$ and from $t_0$ to $t^*$.

\subsection{Dual-Branch Feature Extraction}

The transition to an extreme event in the VCSEL involves two physically
distinct stages: a slow, largely unobservable accumulation of energy in the
suppressed polarization mode, followed by a sudden nonlinear burst triggered by
Langevin noise~\cite{Racca2021}. The two branches are designed to capture
these stages separately. The RC branch integrates the slow precursor drift over
long timescales through its recurrent dynamics. The NGRC branch acts as an
instantaneous high-order nonlinear filter, sensitive to the rapid geometric
sharpening of the burst trajectory in phase space.

\subsubsection{Reservoir Computing Branch}

The RC branch maps the scalar input $x_t$ into a nonlinear state via a
leaky-integrator echo state network:
\begin{equation}
\mathbf{r}_t = (1-\alpha_r)\mathbf{r}_{t-1}
  + \alpha_r\tanh\bigl(\mathbf{W}_{\mathrm{in}}x_t
  + \mathbf{W}\mathbf{r}_{t-1}\bigr),
\label{eq:rc}
\end{equation}
where $\mathbf{r}_t \in \mathbb{R}^{N_r}$ is the reservoir state,
$\mathbf{W}_{\mathrm{in}} \in \mathbb{R}^{N_r \times 1}$ the input weight
matrix, and $\mathbf{W} \in \mathbb{R}^{N_r \times N_r}$ the sparse recurrent
weight matrix. Both matrices are randomly initialized and held fixed during
training. The recurrent matrix $\mathbf{W}$ has spectral radius $\rho$ and is
scaled accordingly. The leaking rate $\alpha_r$ controls the effective memory
timescale.

A key advantage of the hybrid architecture is that the RC branch is relieved
of fitting sharp transients, a task explicitly delegated to the NGRC branch.
This allows the reservoir to function purely as a low-dimensional aggregator of
temporal memory, which in turn permits a much smaller reservoir size of
$N_r = 100$ compared to the large monolithic reservoirs typically required for
highly nonlinear excursions~\cite{Wang2024, Vlachas2020}.

\subsubsection{Next-Generation Reservoir Computing Branch}

The NGRC branch constructs a deterministic polynomial feature vector from
delay-embedded observations~\cite{Gauthier2021, Chepuri2024, Barbosa2022, Bollt2021}. A
delay-embedded state vector is formed as
\begin{equation}
\mathbf{v}_t = \left[ x_t, x_{t-\tau}, \dots,
  x_{t-(D-1)\tau} \right]^\top \in \mathbb{R}^D,
\end{equation}
with embedding dimension $D$ and delay interval $\tau$. The full NGRC feature
vector $\boldsymbol{\phi}_t$ concatenates a constant bias, the linear vector
$\mathbf{v}_t$, squared terms $v_{t,i}^2$, all unique pairwise products
$v_{t,i}v_{t,j}$ for $i < j$, and local window statistics computed over a
trailing window of $L_w$ steps: the mean, standard deviation, and range.

The inclusion of these rolling statistics deserves comment. The polynomial
cross-terms in NGRC are known to be sensitive to dynamical
noise~\cite{Chepuri2024, Patel2023}, and the SFM dynamics are
driven by strong Langevin forcing. Direct polynomial embedding under these
conditions could amplify noise severely. The local statistics act as a
low-pass filter that suppresses high-frequency stochastic spikes while
preserving the macroscopic burst signature. This works because the envelope of
an extreme event evolves on a nanosecond timescale, whereas the Langevin noise
has a correlation time much shorter than the sampling interval. The local mean,
standard deviation, and range effectively smooth out the fast fluctuations
while retaining the slow-varying burst envelope.

\subsection{Feature Fusion, Classification, and Conditioned Regression}

The fused feature vector at each time step is
\begin{equation}
\mathbf{f}_t = \bigl[\mathbf{r}_t \mid \boldsymbol{\phi}_t\bigr]
  \in \mathbb{R}^{N_r + d_\phi}.
\end{equation}
All features are standardized to zero mean and unit variance before being
passed downstream. This prevents the larger NGRC branch from
disproportionately dominating the ridge penalty.

\textbf{Stage~1 (Detection).}
A ridge classifier is trained on $\{(\mathbf{f}_t, y_t)\}$. The continuous
classifier score is
\begin{equation}
s_t = \mathbf{w}_c^\top \mathbf{f}_t + b_c,
\end{equation}
where the weight vector $\mathbf{w}_c$ and bias $b_c$ are obtained by
minimizing the regularized loss
\begin{equation}
\mathcal{L}_c = \sum_{t}(y_t - s_t)^2 + \lambda_c\|\mathbf{w}_c\|^2,
\end{equation}
with $\lambda_c$ being the regularization coefficient. The decision threshold
$\delta^*$ is selected on the validation set to maximize the Symmetric Extremal
Dependence Index (SEDI):
\begin{equation}
\mathrm{SEDI} = \frac{\ln F - \ln H - \ln(1{-}F) + \ln(1{-}H)}%
{\ln F + \ln H + \ln(1{-}F) + \ln(1{-}H)},
\label{eq:sedi}
\end{equation}
where $H = \mathrm{TP}/(\mathrm{TP}+\mathrm{FN})$ is the hit rate and
$F = \mathrm{FP}/(\mathrm{FP}+\mathrm{TN})$ the false-alarm rate. SEDI is
designed specifically for rare-event evaluation: it penalizes missed detections
and false alarms jointly on a symmetric logarithmic scale, remaining well
calibrated even when the event base rate is very low~\cite{Ferro2011}.

\textbf{Event-level aggregation.} Point-wise classifier scores are converted to
event-level decisions by window aggregation. For each threshold-crossing event,
the shifted warning window yields one aggregated score,
$s_e = \max_{t \in [t_0 - W - \Delta W, t_0 - \Delta W]} s_t$. Here,
$\Delta W$ shifts the warning window relative to the event onset but does not
change its duration, which remains $W$. Contiguous non-event regions, defined
as any interval that does not overlap a warning window or a post-event
relaxation tail, are tiled with non-overlapping sub-windows of the same length
$W$; each sub-window contributes one aggregated score,
$s_n = \max_{t \in \mathcal{W}_n} s_t$, where $\mathcal{W}_n$ denotes a
non-event window of length $W$. Thus, event and non-event samples use the same
window length and the same maximum-score aggregation rule. The hit rate $H$ and false-alarm rate $F$ in
Eq.~\eqref{eq:sedi} are computed at the sampling-point level over the full test
sequence. All points inside the warning windows are treated as positive
samples, while the remaining eligible points are treated as negative samples.
Therefore, the resulting SEDI measures point-wise precursor detection.

\begin{figure}[t]
\centering
\includegraphics[width=\linewidth]{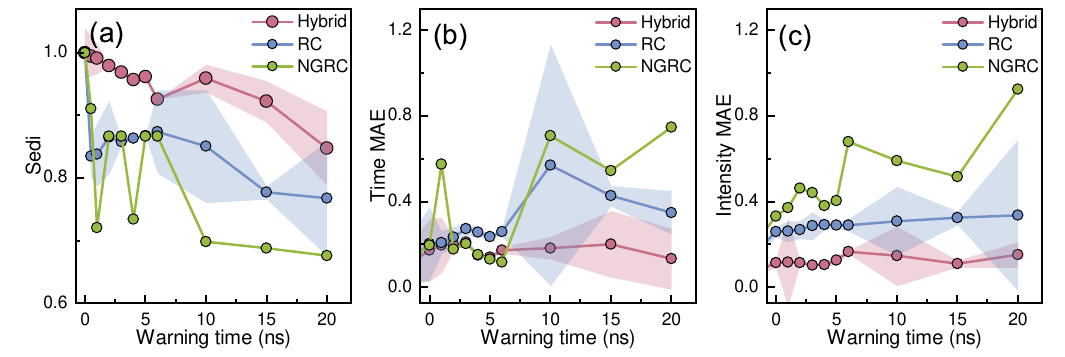}
\caption{\textbf{Prediction performance as a function of warning time
$\Delta W \cdot \Delta t_s$}, representing the forward shift of the warning
window prior to the threshold-crossing index $t_0$. (a) SEDI. (b)
Time-to-peak MAE. (c) Peak intensity MAE. The hybrid RC--NGRC framework
(pink) is compared against the RC-only (blue) and NGRC-only (green)
ablations. Results show the mean across five independent random seeds.
The warning time on the horizontal axis indicates the minimum advance
before $t_0$, not the full alarm-to-peak interval.}
\label{fig:lineplot}
\end{figure}

\textbf{Stage~2 (Conditioned Prediction).}
When an alarm fires ($s_{t_{\mathrm{fa}}} \geq \delta^*$), the feature
snapshot $\mathbf{f}_{t_{\mathrm{fa}}}$ is passed to two independent ridge
regressors:
\begin{align}
\widehat{\Delta t} &= \mathbf{w}_{\Delta t}^\top \mathbf{f}_{t_{\mathrm{fa}}}
  + b_{\Delta t}, \\
\widehat{I}^* &= \mathbf{w}_{I^*}^\top \mathbf{f}_{t_{\mathrm{fa}}}
  + b_{I^*}.
\end{align}
These regressors are trained exclusively on the set of true-positive alarm
snapshots $\mathcal{S}_{\mathrm{TP}}$ from the training set. To construct
$\mathcal{S}_{\mathrm{TP}}$, we first train the Stage~1 classifier on the
training data, then apply it to the same partition and retain only those time
steps where the classifier correctly issued an alarm and an event indeed follows
within the warning window. The weights are determined by minimizing
\begin{align}
\mathcal{L}_{\Delta t} &= \sum_{t \in \mathcal{S}_{\mathrm{TP}}}
  (\Delta t - \widehat{\Delta t})^2
  + \lambda_{\Delta t}\|\mathbf{w}_{\Delta t}\|^2, \\
\mathcal{L}_{I^*} &= \sum_{t \in \mathcal{S}_{\mathrm{TP}}}
  (I^* - \widehat{I}^*)^2
  + \lambda_{I^*}\|\mathbf{w}_{I^*}\|^2.
\end{align}
Training on $\mathcal{S}_{\mathrm{TP}}$ converts the original heavily
imbalanced time series into a near-homogeneous regression task focused entirely
on verified precursor states.

\subsection{Hyperparameter Selection}

All hyperparameters are optimized using Bayesian optimization on the training
partition with 5-fold time-series cross-validation to prevent data leakage. The
objective for the classifier is to maximize the mean SEDI over the validation
folds, whereas the objectives for the regressors are to minimize the MAE
calculated from the true-positive snapshots. The search ranges for the RC
hyperparameters are spectral radius $\rho \in [0.1, 0.99]$, input scaling
$\sigma_{\text{in}} \in [0.1, 3.0]$, and leaking rate
$\alpha_r \in [0.05, 1.0]$. The regularization coefficients
$\lambda_c$, $\lambda_{\Delta t}$, and $\lambda_{I^*}$ are optimized in the
logarithmic range $[10^{-2}, 10^3]$. The optimal values reported in
Table~\ref{tab:all_params} correspond to the configuration achieving the
highest average validation SEDI for the classifier and the lowest average
validation MAE for the regressors. The selected hyperparameters are fixed and
then used for all five independent random seeds reported in
Section~\ref{sec:results}.

\section{Results}
\label{sec:results}

All models are trained on the $6\,\mu\mathrm{s}$ training partition and
evaluated on the disjoint $2\,\mu\mathrm{s}$ test set. The training set
contains approximately 179 threshold-crossing events and the test set about 42
events, providing reasonable statistics for rare-event evaluation. We report
mean absolute error (MAE) as the primary regression metric because MAE is less
sensitive to outliers than root mean square error (RMSE), which is advantageous
when evaluating sparse large deviations that may occur in extreme event
forecasting. Three configurations sharing the same two-stage pipeline are
compared: the full hybrid RC--NGRC architecture, an RC-only ablation with the
NGRC branch excluded, and an NGRC-only ablation with the RC branch excluded.
Metrics are averaged over five independent random seeds. The horizontal axis in
all plots is the warning time $\Delta W \cdot \Delta t_s$, the forward shift of
the warning window defined in Section~\ref{sec:ee_def}.

\begin{figure}[t]
    \centering
    \includegraphics[width=\textwidth]{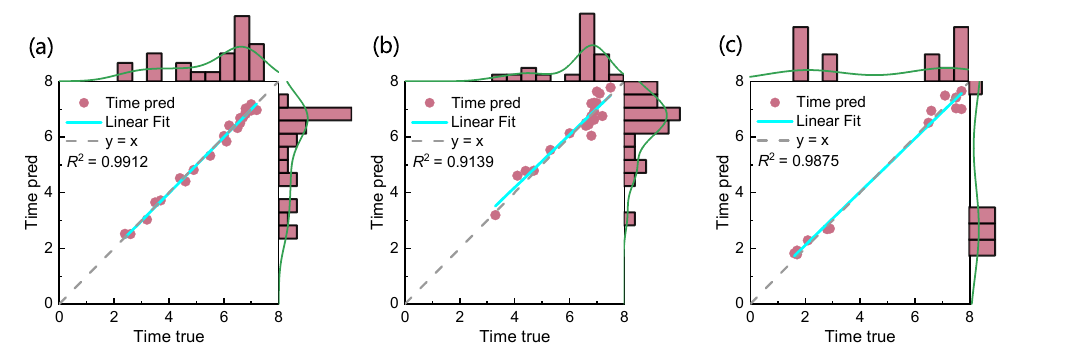}
    \vspace{1.5em}
    \includegraphics[width=\textwidth]{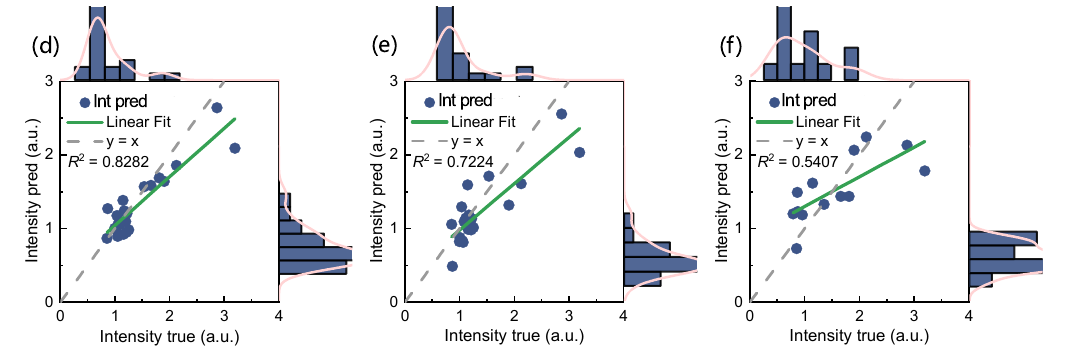}
    \caption{\textbf{Scatter plots comparing predicted and ground truth values
    for time to peak (a, b, c) and peak intensity (d, e, f)} at representative
    warning times. The columns correspond to the hybrid RC--NGRC framework
    (a, d), the RC-only ablation (b, e), and the NGRC-only ablation (c, f).
    The dashed diagonal represents the ideal $y=x$ relationship, while the
    solid line indicates the linear regression fit. Marginal histograms display
    the distributions of the ground truth (top of each panel) and predicted
    (right of each panel) values. Results are shown for the random seed
    yielding the median SEDI performance.}
    \label{fig:scatter}
\end{figure}

\subsection{Main Framework Performance}

Fig.~\ref{fig:lineplot} summarizes the effect of moving the warning window
progressively farther from the threshold-crossing time. The shaded bands represent $\pm$1 standard deviation over five independent random seeds, quantifying the variability due to reservoir initialization. Detection remains
reliable throughout the tested range: the hybrid model retains a SEDI above
$0.8$ and is more stable than either single-branch model at long warning times.
Between $10$ and $20$~ns, its SEDI decreases from $0.9590$ to $0.8475$, an
average reduction of $1.12\times10^{-2}$ per nanosecond. Most of this loss occurs
beyond $15$~ns, where SEDI falls from $0.9220$ to $0.8475$. Moving the warning
window this far from the event therefore mainly reduces the detectability of
the precursor, as the measured trajectory contains less information about the
subsequent threshold crossing.

The regression results show a weaker dependence on warning time. Across the
full range, the hybrid time-to-peak MAE varies from $0.1323$ to $0.2102$~ns,
while the peak-intensity MAE remains between $0.1024$ and $0.1647$~a.u. Beyond
$10$~ns, the timing error shows no systematic increase, and the intensity MAE
changes only from $0.1457$ to $0.1520$~a.u. This distinction is significant:
increasing the lead time makes it harder to issue a correct alarm, but once a valid precursor has been detected, the conditioned regressors retain similar quantitative accuracy. The RC-only model gradually loses timing precision,
whereas the NGRC-only model deteriorates more strongly in both detection and
intensity prediction, consistent with the complementary roles of recurrent
memory and local waveform geometry.

The scatter plots in Fig.~\ref{fig:scatter} offer a more granular view. For
the hybrid model, predictions cluster tightly around the ideal diagonal for
both targets. The linear fit nearly coincides with the $y = x$ line, indicating
negligible systematic bias. Marginal distributions of predicted and
ground-truth values are well matched, which means the model reproduces the full
dynamic range of the event population rather than collapsing toward the mean.
This behavior holds across all five seeds.

\begin{figure}[t]
    \centering
    \includegraphics[width=\linewidth]{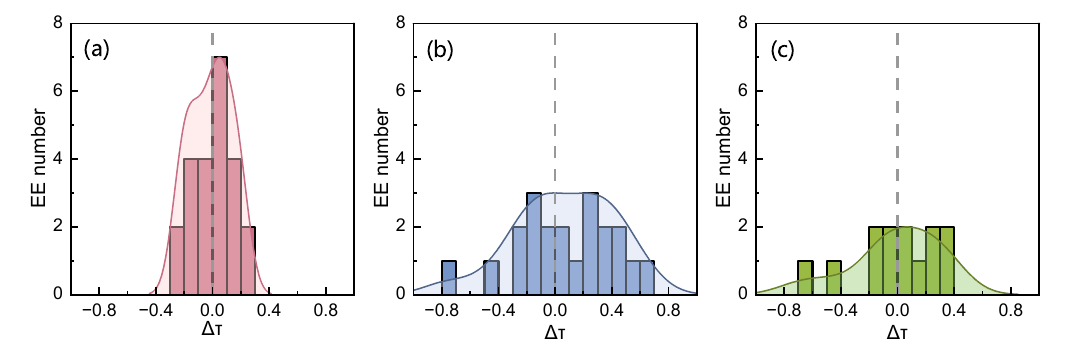}
    \vspace*{1.5em}
    \includegraphics[width=\linewidth]{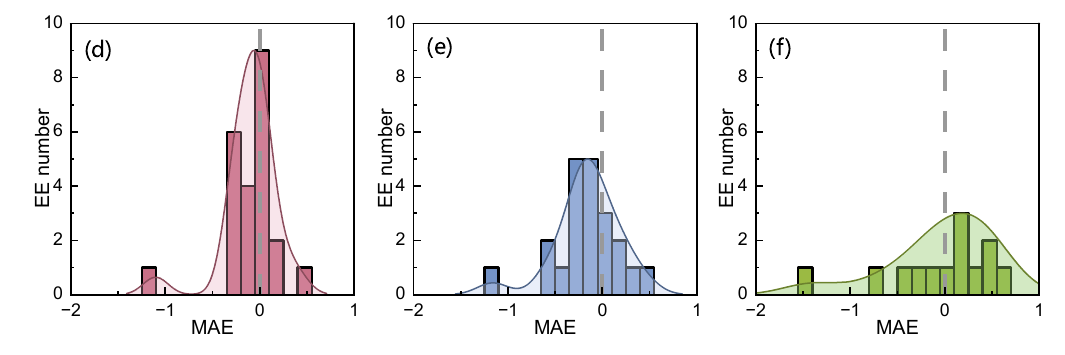}
    \caption{\textbf{Prediction error distributions for time to peak (a, b, c)
    and peak intensity (d, e, f).} The columns correspond to the hybrid
    RC--NGRC framework (a, d; pink), the RC-only ablation (b, e; blue), and
    the NGRC-only ablation (c, f; green). Bars represent normalized histograms,
    solid curves denote kernel density estimates, and dashed vertical lines
    indicate zero prediction error.}
    \label{fig:errdist}
\end{figure}

Error distributions for the hybrid configuration in Fig.~\ref{fig:errdist}
are sharply peaked near zero and roughly symmetric, with tails that decay
rapidly. The near-zero mean confirms that the regressors are unbiased: they
neither systematically overestimate nor underestimate time-to-peak or peak
intensity. The statistical properties of these error distributions, including mean, standard deviation, skewness, and kurtosis, are presented in Table 1 for all three configurations.

\subsection{Ablation Study: Dual-Branch Feature Fusion}

The RC-only and NGRC-only ablations isolate what each branch contributes. Both
are evaluated on the same test set under the same five seeds.

\textbf{RC-only ablation.}
The RC-only variant maintains acceptable SEDI across most of the warning-time
range, but its time-to-peak MAE spikes at intermediate shifts, reaching roughly
three times the hybrid value. The scatter plots tell the same story. Prediction
spread widens for time-to-peak, and the regression fit rotates away from the
diagonal: early-arriving events are underestimated, later ones overestimated.
Error distributions broaden and develop heavier tails. The pattern is
physically interpretable. Without the NGRC branch, the system has no access to
short-term waveform features like local curvature and inter-sample increments.
These carry the fine-grained information needed to resolve the stochastic
Langevin-driven kicks that determine the exact peak timing.

\textbf{NGRC-only ablation.}
The NGRC-only configuration degrades on detection and regression at the same
time. SEDI drops substantially at intermediate warning times, which is
expected: the NGRC branch has no recurrent memory and cannot accumulate the
slow precursor drift that builds over many time steps before an event. The
instantaneous nonlinear features alone are insufficient to reliably distinguish
a genuine burst from high-frequency background noise. Intensity MAE climbs
steeply and monotonically with warning time. The scatter plots show the worst
degradation of the three configurations, with predicted intensity values
compressed toward the population mean and the regression fit displaced well
below the diagonal.

\definecolor{revYellow}{RGB}{255,248,190}
\definecolor{revYellowHead}{RGB}{255,238,130}
\definecolor{revYellowGroup}{RGB}{255,243,165}
\newcommand{\RevTableWidth}{0.88\textwidth}

Taken together, the ablations indicate that the two feature branches are not interchangeable. The recurrent state is primarily responsible for maintaining reliable detection over long contexts, whereas the NGRC features improve peak localization and preserve the intensity range. Their fusion is therefore necessary to achieve balanced performance across detection and both regression targets.

The event-level statistics in Table~\ref{tab:error_statistics_regression_data} further quantify this pattern. Timing errors for the hybrid model are approximately centered near zero, featuring a mean of $-0.0111$~ns and a standard deviation of $0.1428$~ns (compared with $0.3373$~ns for RC-only and $0.2994$~ns for NGRC-only). Conversely, intensity errors exhibit a moderate negative bias (mean of $-0.0984$~a.u.) and a pronounced left tail, indicating the occasional underestimation of high-intensity events. This negative skew may be explained as: the most extreme bursts are driven by rare, strong Langevin kicks that produce waveforms with sharper curvature, which are harder to reconstruct from the limited number of training examples. Although the intensity errors retain this longer negative tail, the fused representation substantially narrows the timing-error distribution and reduces overall regression dispersion.

\begin{table*}[t]
\centering
\caption{Statistical characteristics of prediction errors for the hybrid model and ablation baselines.}
\label{tab:error_statistics_regression_data}
\scriptsize
\renewcommand{\arraystretch}{1.15}
\setlength{\tabcolsep}{4.5pt}
\begin{tabular*}{0.98\textwidth}{@{\extracolsep{\fill}}llrrrrrrrrrr@{}}
\toprule
\textbf{Model}
& \textbf{Error}
& \textbf{Mean}
& \textbf{Std.}
& \textbf{Skew.}
& \textbf{Excess kurt.}
& \textbf{MAE}
& \textbf{RMSE}
& \textbf{Median}
& \textbf{Min}
& \textbf{Max} \\
\midrule

Hybrid
& $e_{\Delta t}$ (ns)
& -0.0111
& 0.1428
& -0.0802
& -1.1490
& 0.1218
& 0.1432
& 0.0092
& -0.2582
& 0.2214 \\

Hybrid
& $e_I$ (a.u.)
& -0.0984
& 0.2702
& -1.8536
& 5.9092
& 0.1774
& 0.2876
& -0.0306
& -1.1006
& 0.4018 \\

RC-only
& $e_{\Delta t}$ (ns)
& 0.0647
& 0.3373
& -0.4488
& -0.2194
& 0.2851
& 0.3434
& 0.0926
& -0.7585
& 0.6307 \\

RC-only
& $e_I$ (a.u.)
& -0.1625
& 0.3304
& -0.9530
& 1.9722
& 0.2701
& 0.3682
& -0.1699
& -1.1601
& 0.4484 \\

NGRC-only
& $e_{\Delta t}$ (ns)
& -0.0114
& 0.2994
& -0.7976
& -0.0236
& 0.2343
& 0.2996
& 0.0078
& -0.6941
& 0.3877 \\

NGRC-only
& $e_I$ (a.u.)
& -0.0398
& 0.5348
& -1.1845
& 0.8839
& 0.4047
& 0.5363
& 0.1193
& -1.4122
& 0.6228 \\

\bottomrule
\end{tabular*}

\vspace{0.6ex}
\begin{minipage}{0.94\textwidth}
\footnotesize
\textit{Note:} The timing and intensity errors are defined as
$e_{\Delta t}=\widehat{\Delta t}-\Delta t$ and
$e_I=\widehat I^{*}-I^{*}$, respectively.
The timing error is reported in nanoseconds. 
\end{minipage}
\end{table*}

\subsection{Sensitivity, Internal Ablation, and Computational Cost}

The temporal sensitivity results in Table~\ref{tab:sensitivity_temporal} show
that the method is stable over a moderate range of analysis windows. SEDI
remains high for warning-window lengths between 3.0 and 5.0~ns, whereas longer
windows produce a clear increase in timing error. The best overall result is
obtained with a delay interval of 0.1~ns, suggesting that the short-term
waveform geometry is most effectively resolved at the native sampling
interval. Performance is relatively stable for $L_w=1.5$--$2.5$~ns. This
interval is comparable to the characteristic rising-edge duration of
the simulated extreme pulses. A window
within this range is therefore sufficiently long to capture the
macroscopic growth of the burst, while avoiding excessive averaging
over earlier background dynamics.

\begin{table*}[t]
\centering
\caption{Sensitivity analysis of the main temporal hyperparameters.}
\label{tab:sensitivity_temporal}
\scriptsize
\renewcommand{\arraystretch}{1.15}
\setlength{\tabcolsep}{4pt}
\begin{tabular*}{\RevTableWidth}{@{\extracolsep{\fill}}llccc@{}}
\toprule
\textbf{Parameter}
& \textbf{Setting}
& \textbf{SEDI}
& \textbf{Time MAE (ns)}
& \textbf{Intensity MAE} \\
\midrule

\multicolumn{5}{@{}l}{\textbf{Warning-window length $W$}} \\
$W$ & 3.0 ns & $0.9363$ & 0.166 & 0.171 \\
$W$ & 4.0 ns & $0.9383$ & 0.181 & 0.140 \\
$W$ & 5.0 ns & $0.9253$ & 0.173 & 0.165 \\
$W$ & 6.0 ns & $0.8925$ & 0.337 & 0.312 \\
$W$ & 7.0 ns & $0.9055$ & 0.360 & 0.299 \\

\midrule
\multicolumn{5}{@{}l}{\textbf{Delay interval $\tau$}} \\
$\tau$ & 0.1 ns & $0.9253$ & 0.173 & 0.165 \\
$\tau$ & 0.2 ns & $0.8714$ & 0.419 & 0.249 \\
$\tau$ & 0.3 ns & $0.8736$ & 0.415 & 0.205 \\

\midrule
\multicolumn{5}{@{}l}{\textbf{Local statistics window length $L_w$}} \\
$L_w$ & 1.0 ns & $0.9083$ & 0.145 & 0.179 \\
$L_w$ & 1.5 ns & $0.9259$ & 0.138 & 0.172 \\
$L_w$ & 2.0 ns & $0.9253$ & 0.173 & 0.165 \\
$L_w$ & 2.5 ns & $0.9199$ & 0.117 & 0.168 \\
$L_w$ & 3.0 ns & $0.9031$ & 0.185 & 0.185 \\
\bottomrule
\end{tabular*}
\end{table*}

The finer NGRC ablation in Table~\ref{tab:ngrc_internal_ablation} explains why
the rolling statistics are retained. Removing the local mean, standard
deviation, and range lowers SEDI from $0.9253$ to $0.8469$ and increases both
regression errors, whereas removing individual polynomial components has a
smaller effect. Under Langevin forcing, the local statistics therefore provide
useful envelope information rather than merely increasing the feature
dimension.

\begin{table*}[t]
\centering
\caption{Fine grained ablation analysis of the NGRC internal modules under the default warning setting.}
\label{tab:ngrc_internal_ablation}
\scriptsize
\renewcommand{\arraystretch}{1.15}
\setlength{\tabcolsep}{5pt}
\begin{tabular*}{0.86\textwidth}{@{\extracolsep{\fill}}lrrrr@{}}
\toprule
\textbf{Variant}
& \textbf{SEDI}
& \textbf{Time MAE (ns)}
& \textbf{Intensity MAE}
& \textbf{NGRC dim} \\
\midrule
Full NGRC
& $0.9253$
& 0.173
& 0.165
& 70 \\
Without local statistics
& $0.8469$
& 0.404
& 0.305
& 67 \\
Without interaction terms
& $0.9239$
& 0.194
& 0.178
& 60 \\
Without quadratic terms
& $0.9096$
& 0.167
& 0.172
& 42 \\
Linear only
& $0.9123$
& 0.179
& 0.176
& 29 \\
\bottomrule
\end{tabular*}

\vspace{0.6ex}
\begin{minipage}{0.86\textwidth}
\footnotesize
\end{minipage}
\end{table*}

Finally, Table~\ref{tab:computational_complexity} shows that adding the NGRC
branch does not make the cascade prohibitively large. The hybrid model contains
510 trainable readout parameters and requires 212.7~s for fitting in the
present Python CPU implementation. Its online inference time is
10.28~$\mu$s per sample, compared with 9.28~$\mu$s for RC-only. These timings
are implementation-dependent, but the small difference confirms that the two
Stage~2 outputs add little online overhead because they are obtained from
linear ridge readouts.

\begin{table*}[t]
\centering
\caption{Computational complexity and inference efficiency of different architectures.}
\label{tab:computational_complexity}
\scriptsize
\renewcommand{\arraystretch}{1.15}
\setlength{\tabcolsep}{4pt}
\begin{tabular*}{\RevTableWidth}{@{\extracolsep{\fill}}lrrrr@{}}
\toprule
\textbf{Architecture}
& \textbf{Total Parameters}
& \textbf{Fit Time (s)}
& \textbf{Inference per Sample}
& \textbf{Throughput (samples/s)} \\
\midrule
RC
& 300
& 153.9
& 9.28 $\mu$s
& 107,724 \\

NGRC
& 210
& 49.6
& 0.81 $\mu$s
& 1,227,069 \\

\textbf{Hybrid (Cascade)}
& \textbf{510}
& \textbf{212.7}
& \textbf{10.28 $\mu$s}
& \textbf{97,225} \\
\bottomrule
\end{tabular*}
\end{table*}

\subsection{Comparison with Prior Work}

The closest related study is the dual-training RC method of Wang et
al.~\cite{Wang2024}, which was evaluated on the same simulated laser system.
That study focused on binary early warning and reported point-wise
classification accuracy at selected lead times, but it did not estimate either
the time-to-peak or the peak intensity.

A direct quantitative comparison with the present results is not appropriate
because the two studies use different evaluation protocols. In particular, the
reported accuracy in Wang et al.~\cite{Wang2024} is based on a different sample
construction and is strongly affected by the proportion of non-event samples.
By contrast, the present study aggregates point-wise scores into event-level
samples, assigns one decision to each extreme event or non-event window, and
uses SEDI as the primary detection metric. Accuracy and SEDI therefore measure
different aspects of performance and cannot be converted or compared directly.

Under the event-level evaluation protocol adopted here, the hybrid model
maintains stable detection performance over a broad range of warning-window
shifts and additionally produces quantitative estimates of $\Delta t$ and
$I^*$. The comparison with the previous study is therefore limited to the
methodological scope of the two approaches: the earlier method provides a
binary warning, whereas the present cascade extends the prediction task from
event detection to the estimation of event timing and magnitude. The significant improvement in prediction horizon, from approximately 15$\tau_a$ in the prior work to 125$\tau_a$ in the present framework, further highlights the advantage of the dual-stage design.

\section{Discussion}

Cascaded conditioning decouples the regression input distribution from the
prior over extreme events. Training a single regressor directly on the full
time series forces the estimator to fit two very different populations under a
shared loss: a background distribution $p_{\text{bg}}(x)$ that dominates the
sample count but carries no informative $\Delta t$ or $I^*$ signal, and a
precursor distribution $p_{\text{pre}}(x)$ that is sparse but predictive.
Empirical risk minimization on the mixture
$p(x) = (1-\pi) p_{\text{bg}}(x) + \pi p_{\text{pre}}(x)$ with $\pi < 0.01$
is dominated by the bias-minimizing solution for $p_{\text{bg}}$, which is the
constant prediction of a non-event. The resulting estimator has low variance on
backgrounds and high variance on transients, the opposite of what the task
requires.

Stage~1 acts as a gating operator that restricts the support of the Stage~2
input distribution to the alarm submanifold $S_{\text{TP}}$, which is a sample
approximation of $p_{\text{pre}}$. The regressors are therefore trained and
tested on a single well-localized distribution. Under this conditioning the
bias term collapses because the targets $\Delta t$ and $I^*$ are no longer
diluted by null labels, while the variance term stays bounded because the
precursor snapshots share common geometric structure, as visible in the marginal
ground-truth distributions of Fig.~4(a, d). The zero-centered, roughly
symmetric error distributions in Fig.~5 follow from this bias--variance
redistribution rather than from any resampling, class weighting, or focal loss
heuristic.

One feature snapshot serves both prediction targets. Prior reservoir computing
work has addressed either binary event detection~\cite{Wang2024, Suda2020}
or amplitude statistics~\cite{Racca2021}, but not both from the same model.
The fused RC--NGRC feature vector at the alarm instant encodes enough
information to estimate $\Delta t$ and $I^*$ simultaneously with low error.
The two ridge regressors share this input and are solved in closed form, so the
second target adds essentially no computational cost.

The two branches capture different dynamical timescales. The ablation results
show a clean functional split. Removing the NGRC branch hurts timing precision.
Removing the RC branch collapses detection and intensity prediction together.
The RC integrates information through recurrent dynamics over many time steps,
picking up the slow carrier-density drift that builds before a macroscopic
burst. The NGRC, constructed from polynomial combinations of recent
delay-embedded samples, acts as a high-frequency filter sensitive to the rapid
waveform sharpening driven by stochastic noise. Stable performance requires
both.

\textbf{The prediction horizon exceeds intrinsic timescales by a wide margin.}
The autocorrelation time of the simulated dynamics is
$\tau_a \approx 0.16\,\mathrm{ns}$~\cite{Wang2024}. The framework maintains
SEDI above $0.8$ with the warning window shifted $20\,\mathrm{ns}$ before the
threshold crossing, corresponding to roughly $125\,\tau_a$. The actual
alarm-to-peak interval is often longer still, because the threshold crossing
typically precedes the absolute peak by several nanoseconds. Wang
et al.~\cite{Wang2024} reported a useful prediction limit of about $15\,\tau_a$
for binary detection alone.

Beyond the present VCSEL application, the proposed cascade offers a flexible
framework for quantitative forecasting in a broader class of rare-event
systems. Its two-stage structure can be adapted by defining system-specific
event criteria, precursor representations, warning horizons, and regression
targets. For example, in ocean rogue-wave prediction, the first stage could
identify abnormal wave-growth patterns, while the second stage could estimate
the arrival time and maximum wave height. In atmospheric applications, the same
strategy could be used to detect early signatures of extreme anomalies and to
predict their onset time and severity. Extending the framework to these domains
will require suitable multivariate inputs, domain-specific labeling rules, and
validation on long observational records. Such studies may clarify whether the
separation of precursor detection from event-attribute estimation provides a
general advantage for forecasting rare, high-impact phenomena.

\subsection{Limitations}

The present study should be viewed as a proof-of-concept evaluation under
controlled numerical conditions. All results are obtained from VCSEL
trajectories generated by the spin-flip model within a single operating
regime, and therefore do not yet establish performance on laboratory data.
Experimental recordings may introduce detector-bandwidth limitations,
sampling jitter, saturation, slow parameter drift, and noise characteristics
that are not fully represented by the Langevin forcing used in the simulation.
Although the proposed cascade does not rely on a simulation-specific
architecture, its event threshold, warning-window definition, preprocessing
pipeline, and ridge readouts would need to be recalibrated using experimental
training data. Validation on sufficiently long laboratory recordings is
therefore required before practical generalization can be claimed.

The current results also remain subject to finite-sample uncertainty. The test
partition contains approximately 42 threshold-crossing events, and fewer
samples are available for regression because Stage~2 is evaluated only for
events that are correctly detected by Stage~1. The five-seed evaluation
quantifies variability associated with model initialization, but it does not
fully capture uncertainty arising from the limited number of extreme events or
from changes in the underlying physical regime. Moreover, the sensitivity
analysis considers the temporal parameters $W$, $\tau$, and $L_w$, but does
not cover broader variations in the SFM parameters, modulation conditions, or
event prevalence. Future work should therefore examine longer and independent
trajectories, multiple operating regimes, and experimental measurements to
better characterize rare large-error cases and assess robustness under
distributional change.
\section{Conclusion}

We have presented a dual-stage, alarm-triggered cascade framework for the joint
quantitative forecasting of extreme optical events in a pump-modulated VCSEL.
By fusing a traditional RC branch for long-range memory with an NGRC branch for
short-term geometric features, the system simultaneously predicts both the
time-to-peak and peak intensity from a single alarm instant. Evaluated on
simulated data, the hybrid model achieves SEDI $>$ 0.8 with mean absolute
errors near 0.2~ns and 0.2~a.u., respectively, and maintains stable performance up
to a 20~ns warning horizon. Ablation studies demonstrate that the RC branch is
essential for reliable detection while the NGRC branch is critical for precise
timing. The classify-then-regress design intrinsically resolves class imbalance
by restricting regression training to true-positive snapshots, yielding
unbiased predictions without data resampling. This work advances
reservoir-computing-based extreme-event forecasting from binary alarm issuance
to fully quantitative, dual-objective prediction.

Relative to prior reservoir-computing approaches that have been limited to
binary alarm issuance or aggregate amplitude statistics, the proposed
methodology elevates extreme-event forecasting from a qualitative warning to a
quantitative, dual-objective prediction. Future work will focus on validating the framework against experimental laboratory recordings, where measurement
noise, finite detection bandwidth, and non-stationary operating conditions will
provide a more stringent test of robustness and generalizability.


\begin{table}[htbp]
\centering
\caption{Hyperparameters of the experiment.}
\label{tab:all_params}
\small
\renewcommand{\arraystretch}{1.18}
\setlength{\tabcolsep}{5pt}
\begin{tabular}{@{}p{0.48\columnwidth} c p{0.30\columnwidth}@{}}
\toprule
\textbf{Parameter Description} 
& \textbf{Symbol} 
& \textbf{Value} \\
\midrule

\multicolumn{3}{@{}l}{\textbf{Spin-Flip Model (SFM) physical parameters}} \\
\addlinespace[0.2em]
Field decay rate 
& $\kappa$ 
& 300 ns$^{-1}$ \\
Linewidth enhancement factor 
& $\alpha$ 
& 3 \\
Carrier decay rate 
& $\gamma$ 
& 1 ns$^{-1}$ \\
Spin-flip relaxation rate 
& $\gamma_s$ 
& 50 ns$^{-1}$ \\
Linear dichroism 
& $\gamma_a$ 
& 0.1 ns$^{-1}$ \\
Linear birefringence 
& $\gamma_p$ 
& 10 ns$^{-1}$ \\
Spontaneous emission coefficient 
& $\beta$ 
& $10^{-3}$ \\
DC bias 
& $I_{\mathrm{dc}}$ 
& $6.03\,P_{\mathrm{th}}^{\mathrm{TE}}$ \\
Modulation amplitude 
& $\Delta I$ 
& $5.2\,P_{\mathrm{th}}^{\mathrm{TE}}$ \\
Modulation frequency 
& $f_m$ 
& 600 MHz \\

\midrule
\multicolumn{3}{@{}l}{\textbf{RC branch hyperparameters}} \\
\addlinespace[0.2em]
Reservoir size 
& $N_r$ 
& 100 \\
Spectral radius 
& $\rho$ 
& Optimized in $[0.1,\,0.99]$ \\
Input scaling 
& $\sigma_{\mathrm{in}}$ 
& Optimized in $[0.1,\,3.0]$ \\
Leaking rate 
& $\alpha_r$ 
& Optimized in $[0.05,\,1.0]$ \\

\midrule
\multicolumn{3}{@{}l}{\textbf{NGRC branch hyperparameters}} \\
\addlinespace[0.2em]
Embedding dimension 
& $D$ 
& 28 \\
Delay interval 
& $\tau$ 
& 0.1 ns \\
Statistical window length 
& $L_w$ 
& 2.0 ns \\

\midrule
\multicolumn{3}{@{}l}{\textbf{Detection and regression framework settings}} \\
\addlinespace[0.2em]
Threshold multiplier 
& $k$ 
& 8.0 \\
Warning-window length 
& $W$ 
& 5.0 ns \\
Integration step size 
& $\Delta t_i$ 
& 2 ps \\
Sampling interval 
& $\Delta t_s$ 
& 0.1 ns \\
Classifier regularization 
& $\lambda_c$ 
& Optimized in $[10^{-2},\,10^3]$ \\
Time regressor regularization 
& $\lambda_{\Delta t}$ 
& Optimized in $[10^{-2},\,10^3]$ \\
Intensity regressor regularization 
& $\lambda_{I^*}$ 
& Optimized in $[10^{-2},\,10^3]$ \\
\bottomrule
\end{tabular}

\vspace{0.8ex}
\begin{minipage}{0.95\columnwidth}
\footnotesize
\textit{Note:} The optimal hyperparameters are
$\rho = 0.7546$, $\sigma_{\mathrm{in}} = 1.5279$, $\alpha_r = 0.2121$,
$\lambda_c = 84.76$, $\lambda_{\Delta t} = 13.08$, and
$\lambda_{I^*} = 206.78$.
\end{minipage}
\end{table}
\section*{Acknowledgments}

This work was partially supported by the National Natural Science Foundation of
China (Grant Nos.\ 62475206 and 61804036), and the Key Research and Development
Plan of Shaanxi Province, China (Grant No.\ 2024GH-ZDXM-42).

\section*{Conflict of Interest}

There are no conflicts to declare.


\bibliographystyle{elsarticle-num}   
\bibliography{ref}

\begin{thebibliography}{10}
\expandafter\ifx\csname url\endcsname\relax
  \def\url#1{\texttt{#1}}\fi
\expandafter\ifx\csname urlprefix\endcsname\relax\def\urlprefix{URL }\fi
\expandafter\ifx\csname href\endcsname\relax
  \def\href#1#2{#2} \def\path#1{#1}\fi

\bibitem{Dysthe2008}
K.~Dysthe, H.~E. Krogstad, P.~M\"{u}ller, Oceanic rogue waves, Annu. Rev. Fluid
  Mech. 40 (2008) 287--310.
\newblock \href {https://doi.org/10.1146/annurev.fluid.40.111406.102203}
  {\path{doi:10.1146/annurev.fluid.40.111406.102203}}.

\bibitem{Pelinovsky2008}
E.~Pelinovsky, C.~Kharif (Eds.), Extreme Ocean Waves, Springer, Berlin, 2008.

\bibitem{Osborne2010}
A.~R. Osborne, Nonlinear Ocean Waves and the Inverse Scattering Transform,
  Academic Press, Burlington, MA, 2010.

\bibitem{Hudson1991}
H.~S. Hudson, Solar flares, microflares, nanoflares, and coronal heating, Sol.
  Phys. 133 (1991) 357--369.
\newblock \href {https://doi.org/10.1007/BF00149894}
  {\path{doi:10.1007/BF00149894}}.

\bibitem{Stenflo2009}
P.~K. Shukla, L.~Stenflo, Nonlinear propagation of electromagnetic waves in
  magnetized plasmas, Phys. Rev. A 30 (1984) 2110--2112.
\newblock \href {https://doi.org/10.1103/PhysRevA.30.2110}
  {\path{doi:10.1103/PhysRevA.30.2110}}.

\bibitem{Metayer2014}
S.~Elaskar, E.~d. Río, L.~Gutiérrez~Marcantoni, Analytical evaluation of the
  reinjection probability density function in chaotic intermittency, in: 2020
  IEEE Congreso Bienal de Argentina (ARGENCON), 2020, pp. 1--8.
\newblock \href {https://doi.org/10.1109/ARGENCON49523.2020.9505540}
  {\path{doi:10.1109/ARGENCON49523.2020.9505540}}.

\bibitem{Baronio2012}
F.~Baronio, A.~Degasperis, M.~Conforti, S.~Wabnitz, Solutions of the vector
  nonlinear {Schr\"{o}dinger} equations: Evidence for deterministic rogue
  waves, Phys. Rev. Lett. 109 (2012) 044102.
\newblock \href {https://doi.org/10.1103/PhysRevLett.109.044102}
  {\path{doi:10.1103/PhysRevLett.109.044102}}.

\bibitem{Dudley2009}
J.~M. Dudley, G.~Genty, F.~Dias, B.~Kibler, N.~Akhmediev, Modulation
  instability, {Akhmediev} breathers and continuous wave supercontinuum
  generation, Opt. Express 17 (2009) 21497--21508.
\newblock \href {https://doi.org/10.1364/OE.17.021497}
  {\path{doi:10.1364/OE.17.021497}}.

\bibitem{Bonatto2011}
C.~Bonatto, M.~Feyereisen, S.~Barland, M.~Giudici, C.~Masoller, J.~R.~R. Leite,
  J.~R. Tredicce, Deterministic optical rogue waves, Phys. Rev. Lett. 107
  (2011) 053901.
\newblock \href {https://doi.org/10.1103/PhysRevLett.107.053901}
  {\path{doi:10.1103/PhysRevLett.107.053901}}.

\bibitem{Onorato2013}
M.~Onorato, S.~Residori, U.~Bortolozzo, A.~Montina, F.~T. Arecchi, Rogue waves
  and their generating mechanisms in different physical contexts, Phys. Rep.
  528 (2013) 47--89.
\newblock \href {https://doi.org/10.1016/j.physrep.2013.03.001}
  {\path{doi:10.1016/j.physrep.2013.03.001}}.

\bibitem{Akhmediev2016}
N.~Akhmediev, et~al., Roadmap on optical rogue waves and extreme events, J.
  Opt. 18 (2016) 063001.
\newblock \href {https://doi.org/10.1088/2040-8978/18/6/063001}
  {\path{doi:10.1088/2040-8978/18/6/063001}}.

\bibitem{Chowdhury2022}
S.~Chowdhury, A.~Prasad, J.~Kurths, R.~Ramaswamy, Extreme events in dynamical
  systems and random walkers: A review, Phys. Rep. 966 (2022) 1--116.
\newblock \href {https://doi.org/10.1016/j.physrep.2022.04.001}
  {\path{doi:10.1016/j.physrep.2022.04.001}}.

\bibitem{Farazmand2019}
M.~Farazmand, T.~P. Sapsis, Extreme events: Mechanisms and prediction, Appl.
  Mech. Rev. 71 (2019) 050801.
\newblock \href {https://doi.org/10.1115/1.4042065}
  {\path{doi:10.1115/1.4042065}}.

\bibitem{Ummenhofer2017}
C.~C. Ummenhofer, G.~A. Meehl, Extreme weather and climate events with
  ecological relevance: a review, Philos. Trans. R. Soc. B 372 (2017) 20160135.
\newblock \href {https://doi.org/10.1098/rstb.2016.0135}
  {\path{doi:10.1098/rstb.2016.0135}}.

\bibitem{Raymond2020}
C.~Raymond, R.~M. Horton, J.~Zscheischler, O.~Martius, A.~AghaKouchak,
  J.~Balch, S.~G. Bowen, S.~J. Camargo, J.~Hess, K.~Kornhuber, M.~Oppenheimer,
  A.~C. Ruane, T.~Wahl, K.~White, Understanding and managing connected extreme
  events, Nat. Clim. Chang. 10 (2020) 611--621.
\newblock \href {https://doi.org/10.1038/s41558-020-0790-4}
  {\path{doi:10.1038/s41558-020-0790-4}}.

\bibitem{Pathak2018}
J.~Pathak, B.~Hunt, M.~Girvan, Z.~Lu, E.~Ott, Model-free prediction of large
  spatiotemporally chaotic systems from data: A reservoir computing approach,
  Phys. Rev. Lett. 120 (2018) 024102.
\newblock \href {https://doi.org/10.1103/PhysRevLett.120.024102}
  {\path{doi:10.1103/PhysRevLett.120.024102}}.

\bibitem{Ghosh2021}
S.~Ghosh, A.~Bhattacharya, M.~S. Baptista, Reservoir computing on epidemic
  spreading: A case study on {COVID-19} cases, Phys. Rev. E 104 (2021) 014308.
\newblock \href {https://doi.org/10.1103/PhysRevE.104.014308}
  {\path{doi:10.1103/PhysRevE.104.014308}}.

\bibitem{Tanaka2019}
G.~Tanaka, T.~Yamane, J.~B. H\'{e}roux, R.~Nakane, N.~Kanazawa, S.~Takeda,
  H.~Numata, D.~Nakano, A.~Hirose, Recent advances in physical reservoir
  computing: A review, Neural Netw. 115 (2019) 100--123.
\newblock \href {https://doi.org/10.1016/j.neunet.2019.03.005}
  {\path{doi:10.1016/j.neunet.2019.03.005}}.

\bibitem{Jaeger2001}
H.~Jaeger, The" echo state" approach to analysing and training recurrent neural
  networks-with an erratum note', Bonn, Germany: German National Research
  Center for Information Technology GMD Technical Report 148 (2001).

\bibitem{Jaeger2004}
H.~Jaeger, H.~Haas, Harnessing nonlinearity: Predicting chaotic systems and
  saving energy in wireless communication, Science 304 (2004) 78--80.
\newblock \href {https://doi.org/10.1126/science.1091277}
  {\path{doi:10.1126/science.1091277}}.

\bibitem{Maass2002}
W.~Maass, T.~Natschl\"{a}ger, H.~Markram, Real-time computing without stable
  states: A new framework for neural computation based on perturbations, Neural
  Comput. 14 (2002) 2531--2560.
\newblock \href {https://doi.org/10.1162/089976602760407955}
  {\path{doi:10.1162/089976602760407955}}.

\bibitem{Lukosevicius2009}
M.~Luko\v{s}evi\v{c}ius, H.~Jaeger, Reservoir computing approaches to recurrent
  neural network training, Comput. Sci. Rev. 3 (2009) 127--149.
\newblock \href {https://doi.org/10.1016/j.cosrev.2009.03.005}
  {\path{doi:10.1016/j.cosrev.2009.03.005}}.

\bibitem{Racca2021}
A.~Mahata, S.~L. Kingston, S.~Ghosh, S.~K. Dana, T.~Kapitaniak, Learning
  transitions to extreme events using reservoir computing, Phys. Rev. E 112
  (2025) 054207.
\newblock \href {https://doi.org/10.1103/rr6x-gdvc}
  {\path{doi:10.1103/rr6x-gdvc}}.

\bibitem{Vlachas2020}
P.~R. Vlachas, J.~Pathak, B.~R. Hunt, T.~P. Sapsis, M.~Girvan, E.~Ott,
  P.~Koumoutsakos, Backpropagation algorithms and reservoir computing in
  recurrent neural networks for the forecasting of complex spatiotemporal
  dynamics, Neural Netw. 126 (2020) 191--217.
\newblock \href {https://doi.org/10.1016/j.neunet.2020.02.016}
  {\path{doi:10.1016/j.neunet.2020.02.016}}.

\bibitem{Pei2023}
F.~Iglesias-Suarez, P.~Gentine, B.~Solino-Fernandez, T.~Beucler, M.~Pritchard,
  J.~Runge, V.~Eyring, Causally-informed deep learning to improve climate
  models and projections, Journal of Geophysical Research: Atmospheres 129
  (2024) e2023JD039202.
\newblock \href {https://doi.org/https://doi.org/10.1029/2023JD039202}
  {\path{doi:https://doi.org/10.1029/2023JD039202}}.

\bibitem{Nadiga2021}
B.~T. Nadiga, Reservoir computing as a tool for climate predictability studies,
  J. Adv. Model. Earth Syst. 13 (2021) e2020MS002290.
\newblock \href {https://doi.org/10.1029/2020MS002290}
  {\path{doi:10.1029/2020MS002290}}.

\bibitem{Saha2022}
A.~Saha, U.~Feudel, S.~K. Dana, Predicting the data structure prior to extreme
  events from passive observables using echo state network, Front. Appl. Math.
  Stat. 8 (2022) 955044.
\newblock \href {https://doi.org/10.3389/fams.2022.955044}
  {\path{doi:10.3389/fams.2022.955044}}.

\bibitem{Carroll2018}
A.~Banerjee, J.~Pathak, R.~Roy, J.~G. Restrepo, E.~Ott, Using machine learning
  to assess short term causal dependence and infer network links, Chaos: An
  Interdisciplinary Journal of Nonlinear Science 29 (2019) 121104.
\newblock \href {https://doi.org/10.1063/1.5134845}
  {\path{doi:10.1063/1.5134845}}.

\bibitem{Suda2020}
Y.~Suda, K.~Okuda, Emergence of second coherent regions for breathing chimera
  states, Phys. Rev. E 101 (2020) 062203.
\newblock \href {https://doi.org/10.1103/PhysRevE.101.062203}
  {\path{doi:10.1103/PhysRevE.101.062203}}.

\bibitem{Wang2024}
T.~Wang, H.~Zhou, Q.~Fang, Y.~Han, X.~Guo, Y.~Zhang, C.~Qian, H.~Chen,
  S.~Barland, S.~Xiang, G.~L. Lippi, Reservoir computing-based advance warning
  of extreme events, Chaos, Solitons \& Fractals 181 (2024) 114673.
\newblock \href {https://doi.org/https://doi.org/10.1016/j.chaos.2024.114673}
  {\path{doi:https://doi.org/10.1016/j.chaos.2024.114673}}.

\bibitem{Gauthier2021}
D.~J. Gauthier, E.~Bollt, A.~Griffith, W.~A.~S. Barbosa, Next generation
  reservoir computing, Nat. Commun. 12 (2021) 5564.
\newblock \href {https://doi.org/10.1038/s41467-021-25801-2}
  {\path{doi:10.1038/s41467-021-25801-2}}.

\bibitem{Chepuri2024}
R.~Chepuri, D.~Amzalag, T.~M. Antonsen, M.~Girvan, Hybridizing traditional and
  next-generation reservoir computing to accurately and efficiently forecast
  dynamical systems, Chaos: An Interdisciplinary Journal of Nonlinear Science
  34~(6) (2024) 063114.
\newblock \href {https://doi.org/10.1063/5.0206232}
  {\path{doi:10.1063/5.0206232}}.

\bibitem{Barbosa2022}
L.~Grigoryeva, A.~Hart, J.-P. Ortega, Learning strange attractors with
  reservoir systems, Nonlinearity 36~(9) (2023) 4674.
\newblock \href {https://doi.org/10.1088/1361-6544/ace492}
  {\path{doi:10.1088/1361-6544/ace492}}.

\bibitem{Bollt2021}
E.~Bollt, On explaining the surprising success of reservoir computing
  forecasting of chaos: The universal machine learning dynamical system with
  contrast to {VAR} and {DMD}, Chaos 31 (2021) 013108.
\newblock \href {https://doi.org/10.1063/5.0024890}
  {\path{doi:10.1063/5.0024890}}.

\bibitem{wang2025extreme}
S.-Y. Wang, S.-C. Chan, Extreme events in detuned semiconductor lasers under
  delayed mutual injection, Phys. Rev. A 112 (2025) 033505.
\newblock \href {https://doi.org/10.1103/8rrz-5kdf}
  {\path{doi:10.1103/8rrz-5kdf}}.

\bibitem{SanMiguel1995}
M.~San~Miguel, Q.~Feng, J.~V. Moloney, Light-polarization dynamics in
  surface-emitting semiconductor lasers, Phys. Rev. A 52 (1995) 1728--1739.
\newblock \href {https://doi.org/10.1103/PhysRevA.52.1728}
  {\path{doi:10.1103/PhysRevA.52.1728}}.

\bibitem{MartinRegalado1997}
J.~Martin-Regalado, F.~Prati, M.~San~Miguel, N.~B. Abraham, Polarization
  properties of vertical-cavity surface-emitting lasers, IEEE J. Quantum
  Electron. 33 (1997) 765--783.
\newblock \href {https://doi.org/10.1109/3.572151}
  {\path{doi:10.1109/3.572151}}.

\bibitem{Honeycutt1992}
R.~L. Honeycutt, Stochastic {Runge-Kutta} algorithms. {I}. white noise, Phys.
  Rev. A 45 (1992) 600--603.
\newblock \href {https://doi.org/10.1103/PhysRevA.45.600}
  {\path{doi:10.1103/PhysRevA.45.600}}.

\bibitem{KloedenPlaten1992}
P.~E. Kloeden, E.~Platen, Numerical Solution of Stochastic Differential
  Equations, Springer, Berlin, 1992.

\bibitem{Patel2023}
Y.~Zhang, S.~P. Cornelius, Catch-22s of reservoir computing, Phys. Rev. Res. 5
  (2023) 033213.
\newblock \href {https://doi.org/10.1103/PhysRevResearch.5.033213}
  {\path{doi:10.1103/PhysRevResearch.5.033213}}.

\bibitem{Ferro2011}
C.~A.~T. Ferro, J.~Stephenson, Extremal dependence indices: Improved
  verification measures for deterministic forecasts of rare binary events,
  Weather Forecast. 26 (2011) 699--713.
\newblock \href {https://doi.org/10.1175/WAF-D-10-05030.1}
  {\path{doi:10.1175/WAF-D-10-05030.1}}.

\end{thebibliography}

\end{document}